\documentclass{article}
\usepackage{spconf,amsmath,graphicx}
\usepackage{cite}
\usepackage[english]{babel}
\usepackage{epstopdf}

\title{Linear networks based speaker adaptation for Speech Synthesis}
\name{Zhiying Huang, Heng Lu, Ming Lei, Zhijie Yan}
\address{Alibaba Inc., Beijing, China\\
	\{zhiying.hzy, h.lu, lm86501, zhijie.yzj\}@alibaba-inc.com}
\begin{document}
	\ninept
	\maketitle
	\begin{abstract}
		Speaker adaptation methods aim to create fair quality synthesis speech voice font for target speakers while only limited resources available. Recently, as deep neural networks based statistical parametric speech synthesis (SPSS) methods become dominant in SPSS TTS back-end modeling, speaker adaptation under the neural network based SPSS framework has also became an important task. In this paper, linear networks (LN) is inserted in multiple neural network layers and fine-tuned together with output layer for best speaker adaptation performance. When adaptation data is extremely small, the low-rank plus diagonal(LRPD) decomposition for LN is employed to make the adapted voice more stable. Speaker adaptation experiments are conducted under a range of adaptation utterances numbers. Moreover, speaker adaptation from 1) female to female, 2) male to female and 3) female to male are investigated. Objective measurement and subjective tests show that LN with LRPD decomposition performs most stable when adaptation data is extremely limited, and our best speaker adaptation (SA) model with only 200 adaptation utterances achieves comparable quality with speaker dependent (SD) model trained with 1000 utterances, in both naturalness and similarity to target speaker.
	\end{abstract}
	
	\section{Introduction}
	\label{sec:intro}
	% add speaker style
	Given adequate amount of training data from target speaker, one can always build SD acoustic model that generates speech very similar to the target speaker himself or herself. Unfortunately, for the most of time, getting enough data from target speaker is not a trivial task. And building SD voices with limited data and bad phoneme coverage could lead to really poor voice quality and intelligibility. By reusing the information from other existing source speaker models, speaker adaptation can obtain satisfactory voice font for target speaker using only limited target speaker data. In this way, speaker adaptation saves the labor of massive recording, manually transcription and checking, and finally makes the cost of creating new voices small and acceptable.
	
	% HMM-based adaptation
	In conventional Hidden Markov Model (HMM) based speech synthesis system, most adaptation methods firstly build average voice using multiple speakers' data, and then conduct speaker adaptation from the huge average model with small amount target speaker data~\cite{yamagishi2003training}. Compared to the large data requirement for building SD model, speaker adaptation can adapt from speaker independent (SI) model with much smaller amount of target speaker data. Many effective speaker adaptation methods have been proposed under the HMM-based speech synthesis framework. The Maximum Likelihood Linear Regression (MLLR) is originally developed in automatic speech recognition tasks~\cite{leggetter1995maximum}, and it was extended to speech synthesis in~\cite{tamura1998speaker, tamura2001adaptation}. In MLLR, the mean vectors of HMM state distributions of the average voice model are transformed to target speaker dependent model through linear transformation. The Hidden Semi-Markov Model (HSMM) based MLLR adaptation~\cite{yamagishi2006hsmm, yamagishi2007average} transforms not only distributions of spectrum and pitch model but also distribution of duration model. Besides, the Constrained Structural Maximum A Posterior Linear Regression (CSMAPLR) adaptation~\cite{nakano2006constrained, yamagishi2009robust, yamagishi2009analysis} simultaneously transforms the mean vectors and co-variance matrices of state output and state duration distributions of the HSMMs in speech synthesis for speaker adaptation.
	
	% NN-based adaptation
	In recent years, neural networks (NN) based speech synthesis dominates back-end acoustic modeling in speech synthesis due to its powerful modeling capacity. It's proved that the NN-based speech synthesis system obtains better voice quality than conventional HMM-based method with the same number of model parameters~\cite{ze2013statistical}. From then on, many research works have been done to investigate speaker adaptation for NN-based speech synthesis. By combining the information from multiple speakers, multi-speaker DNN~\cite{fan2015multi} is proposed. It's assumed that the difference among different speakers can be learned by different output layers. In this way, each speaker owns his or her own output layer but all hidden layers are shared among all speakers. In~\cite{wu2015study}, different levels of speaker adaptation are performed. I-vectors~\cite{dehak2011front} are augmented with linguistic features to represent speaker identity information at input level, and the Learning Hidden Unit Contributions (LHUC)~\cite{swietojanski2014learning} is used to conduct model adaptation at the middle level. Finally, the feature space transformations are implemented at the output layer. Also, in some multi-speaker speech synthesis systems, i-vector and speaker code representing speaker~\cite{zhao2016speaker, luong2017adapting} are combined with linguistic features as input features for the neural network based acoustic model.
	
	% single-speaker adaptation of TTS synthesis
	In this paper, a new linear networks based speaker adaptation approach is proposed in speech synthesis. Inspired by the use of Linear Networks (LN)~\cite{neto1995speaker, gemello2007linear, li2010comparison} based adaptation method in speech recognition tasks, we introduce LN at multiple layers in the speech synthesis structure, and fine-tune them together with output linear layer for speaker adaptation. Moreover, LRPD decomposition was conducted for LN to remove redundant free parameters when adaptation data is very small. Both objective and subjective experiments show that the proposed methods render good adaptation ability in terms of naturalness and similarity to target speaker.
	
	% organization
	The remainder of this paper is organized as follows: section~\ref{sec:adaptation framework} introduces the adaptation framework in this paper, including 1) the multi-task DNN-BLSTM source acoustic model for speech synthesis baseline, 2) describing the framework of LN based speaker adaptation and 3) introducing LRPD decomposition based LN method in detail, section~\ref{sec:exp} evaluates the adaptation methods by using objective measurement and subjective tests, and section~\ref{sec:conclusion} draws conclusions finally.
	
	\section{Adaptation framework}
	\label{sec:adaptation framework}
	
	\begin{figure}[htb]
		\centering
		\centerline{\includegraphics[scale=0.47]{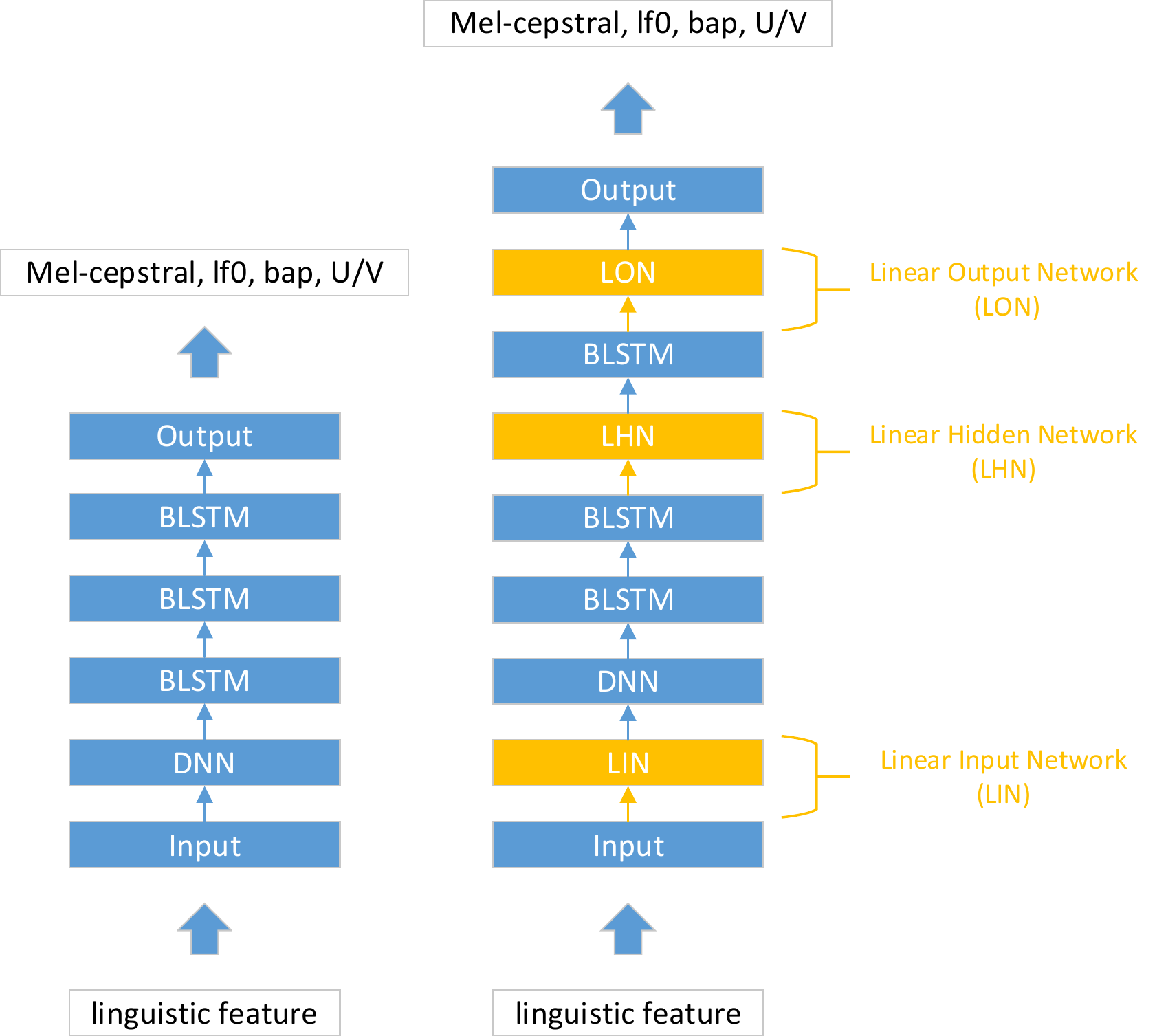}}
		\caption{Multi-task DNN-BLSTM based acoustic model (left) and Linear Network based speaker adaptation (right).}
		\label{fig:LN}
	\end{figure}
	
	\subsection{multi-task DNN-BLSTM based acoustic model}
	\label{subsec:speech_synthesis_system}
	
	Similar to~\cite{fan2014tts, lualibaba}, our back-end base acoustic model is shown on the left side of Fig.~\ref{fig:LN}, which is a multi-task DNN-BLSTM hybrid model. The input features of the back-end base acoustic model are linguistic features, including binary answers to questions about linguistic contexts and numeric values, e.g. absolute positions in different levels of units. The output acoustic features include mel-cepstral coefficients, logarithmic fundamental frequency (log F0) values, band aperiodicities and unvoiced/voiced identity. DNN layer is designed here after input features for better bottom feature extraction, which leads to faster overall convergence. On the left side of Fig.~\ref{fig:LN}, mel-cepstral coefficients, log F0s, band aperiodicities and unvoiced/voiced identity have their separate output layers, while sharing the same lower layers. Our preliminary experiments show that multi-task learning renders more stable synthesis voice than a large single output layer for all acoustic features. The back-end model described in this section serves as source NN structure in the following experiments.
	
	\subsection{Linear Network based adaptation}
	\label{subsec:LN}
	
	LN based adaptation is originally explored in speech recognition tasks, the structure of which is shown on the right side of Fig.~\ref{fig:LN}. LN (the yellow part in Fig.~\ref{fig:LN}) are inserted at multiple layers to source NN-based acoustic model. According to the different positions of LN, LN based adaptation method includes Linear Input Network (LIN~\cite{neto1995speaker}), Linear Hidden Network (LHN~\cite{gemello2007linear}) and Linear Output Network (LON~\cite{li2010comparison}), while LHN can be inserted to any positions between two successive hidden layers. 
	
	When LN is inserted between $l$-th and $l+1$-th hidden layer, the calculation of output of LN $\hat{\textit{\textbf{h}}}^l$ is shown in equation~(\ref{eqn:LN}).
	
	\begin{equation}
	\label{eqn:LN}
	\hat{\textit{\textbf{h}}}^l=\textbf{W}_s\textit{\textbf{h}}^l+\textit{\textbf{b}}_s
	\end{equation}
	Where $\textit{\textbf{h}}^l$ is the activation output at $l$-th hidden layer (or input acoustic features at input layer for LIN), $\textbf{W}_s$ denotes the speaker-specific linear transformation matrix, and $\textit{\textbf{b}}_s$ is the speaker-specific bias vector.
	
	In adaptation process, LN are firstly inserted into specific positions in source model with the linear transformation matrix $\textbf{W}_s$ initializing as identity matrix, and $\textit{\textbf{b}}_s$ is initialized to $0.0$. Then, with adaptation data from target speaker, inserted layers are optimized in back-propagation while keeping other layers fixed. It's worth noting that LIN, LHN and LON can be combined to train a better SA model so as to achieve good adaptation voice quality.
	
	\subsection{LN with Low-rank Plus Diagonal decomposition}
	\label{subsec:LRPD}
	
	\begin{figure}[t]
		\centering
		\centerline{\includegraphics[scale=0.47]{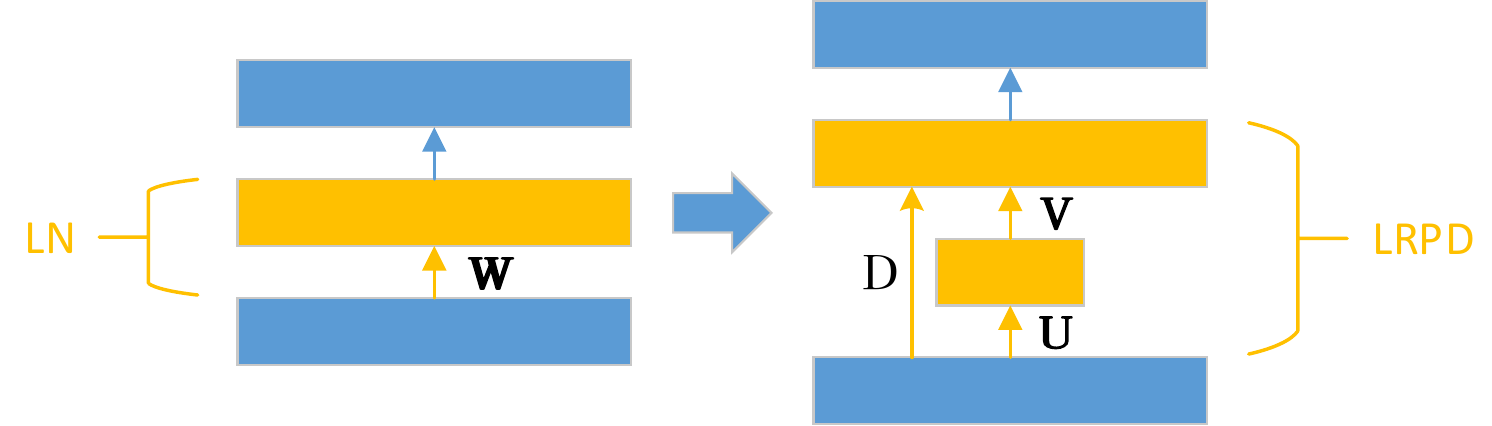}}
		\caption{LN with Low-rank Plus Diagonal decomposition.}
		\label{fig:LRPD}
	\end{figure}
	
	LRPD decomposition is proposed to reduce the number of free adaptation parameters for inserted LN. LN with LRPD decomposition (names as LRPD-LN) can reduce the speaker-specific footprint by 82\% over LN with full adaptation matrix (named as Full-LN) without significant loss of word error rate in speech recognition tasks~\cite{zhao2016low}. The LRPD decomposition restructures the adaptation matrix as a superposition of a diagonal matrix and a low-rank matrix, which is shown in Fig. \ref{fig:LRPD} and equation (\ref{eqn:W}). 
	
	\begin{equation}
	\label{eqn:W}
	\textbf{W}_{s,{k\times k}}\approx\textbf{U}_{s,{k\times r}}\textbf{V}_{s,{r\times k}}+\textbf{D}_{k\times k}
	\end{equation}
	In equation (\ref{eqn:W}), $\textbf{U}_{s,{k\times r}}$ and $\textbf{V}_{s,{r\times k}}$ are two small matrices with dimension $k\times r$ and $r\times k$ respectively. $\textbf{D}_{k\times k}$ is a diagonal matrix. The number of parameters in Full-LN is $k^2$, and the number of parameters in LRPD-LN is $k(2r+1)~(r<<k)$, which is much smaller than that of Full-LN. Moreover, LRPD-LN for speaker adaptation is more suitable than Full-LN when with small target speaker adaptation data, because the number of parameters to be fine-tuned is also smaller.
	
	In paper~\cite{zhao2016low}, the diagonal matrix $\textbf{D}_{k\times k}$ is initialized with identity matrix, which is fine-tuned or fixed during adaptation stage. It's shown that keeping matrix fixed renders comparable performance with the fine-tuning, so we fix diagonal matrix as identity matrix in the following experiments in this work. LRPD-LN can be trained in two ways: 1) initialize two small matrices $\textbf{U}_{s,{k\times r}}$ and $\textbf{V}_{s,{r\times k}}$ randomly, then train all inserted parameters with target speaker data, 2) decompose a well trained Full-LN as the seed model, then fine-tune with target speaker's data. From preliminary experiments, these two training methods shows comparable results, and here for simplicity, we only train LRPD-LN by the first method in the following experiments.

	\section{Experiments}
	\label{sec:exp}
	
	\begin{figure*}[t]
		\centering
		\centerline{\includegraphics[scale=0.395]{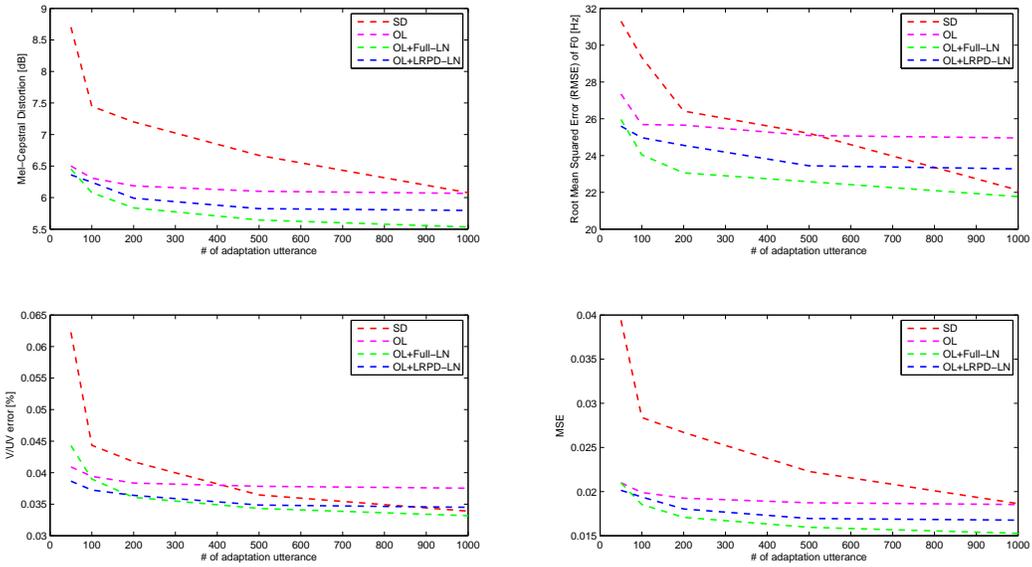}}
		\caption{Objective measurement of validation set utterances (from C-female to B-female).}
		\label{fig:TTS_SAT_xiaoyun_to_xiaoyu_objective}
	\end{figure*}
	
	\subsection{Speech corpus}
	
	Experiments are conducted with a corpus of 3 native Mandarin speakers, all phonetically and prosodically rich. Speech signals are sampled at 16kHz, window size and window shift are 25ms and 5ms in feature extraction, respectively. For all the three speakers in the corpus, one is male (named male-A), and the other two are female (named female-B and female-C). Each speaker has approximately 5000 utterances.
	
	\subsection{Experiment setup}
	For adaptation experiments, adaptation utterances number for each target speaker varies from 50 to 1000. For each target speaker, 200 utterances are held out as validation set for objective measurement, and another 20 utterances are held out as testing set for subjective tests. Here, we only focus on adaptation scenario from one source speaker to one target speaker. The source speaker SD model trained with approximately 5000 utterances is used as source model for speaker adaptation. To investigate adaptation effect between speakers with different distances, three kinds of adaptation experiments are designed: from female to female, from male to female and from female to male. The adaptation from female to female is regarded as adaptation between two similar speakers (easy task), but other two are regarded as adaptation between two far-away speakers (tough task), since two speakers with different genders always have different characters in terms of pitch and spectrum. Moreover, speaker dependent systems of target speaker with different number of utterances are compared as reference in experiments.
	
	Input linguistic features contain 738-dim binary features and 15-dim numerical features, and it's 753-dim totally. Output acoustic feature vectors are 75-dim, comprising 60-dim mel-cepstral coefficients, 3-dim log F0 ($static + \Delta + \Delta\Delta$), 11-dim band aperiodicities and 1-dim UV flag. Linear interpolation of F0 is done over unvoiced segments before modeling, and both input linguistic features and output acoustic features are normalized to 0-mean and unit-variance. In synthesis stage, predicted acoustic features are de-normalized and sent to vocoder WORLD~\cite{morise2016world} for synthesis.
	
	The topology of base DNN-BLSTM model contains a DNN layer of 1024 nodes, and 3 upper layers of BLSTMs with 1024 nodes (512 for each direction). For LRPD-LN, $r$ is set to 10, and the number of parameters in LRPD-LN is $21*k$ while in Full-LN is $k^2$. In training SD model and adaptation model, Minimum Square Error (MSE) is adopted as training criterion and Stochastic Gradient Descent (SGD) based back-propagation is used to optimize model parameters. When training with limited data, we manually tuned learning rate and corresponding hyper-parameters to avoid over-fitting. In informal experiments, it's found that only inserting LN cannot achieve satisfactory adaptation results, and inspired by~\cite{fan2015multi}, we combine fine-tuning output layer with Full-LN/LRPD-LN based method in the following experiments. Different insert positions of LN and different combinations are compared in preliminary experiments, and it's found that only insert LN before last hidden layer and before output layer together would achieve best performance, and this setting will be used in following experiments.
	
	Both objective measurement and subjective tests are conducted in experiments. Objective measurement includes Mel-Cepstral Distortion (MCD), root mean squared error (RMSE) of F0, unvoiced/voiced (U/V) prediction errors and overall MSE in the multi-task output layer. Mean opinion score (MOS) tests on both naturalness and similarity to target speaker are conducted in subjective comparison. Each utterance is listened 5 times by different listeners.
	
	\subsection{Easy task: from female to female}
	\label{subsec:female_to_female}
	
	Adaptation from one female speaker to another female speaker is first evaluated.
	
	\subsubsection{objective measure}
	
	Fig.~\ref{fig:TTS_SAT_xiaoyun_to_xiaoyu_objective} shows the relationship between target speaker adaptation utterances amount and system objective measurement detailed above. "SD" means SD model of target speaker with the same training data as adaptation. "OL" means adaptation with only fine-tuning output layer of source model. "OL + Full-LN" and "OL + LRPD-LN" mean Full-LN and LRPD-LN based adaptation methods and combined with fine-tuning output layer respectively.
	
	As is shown in Fig.~\ref{fig:TTS_SAT_xiaoyun_to_xiaoyu_objective}, with the number of training/adaptation utterances increasing, the objective distance of all systems becomes closer to 0, which indicates all systems perform better with more data. Compared to SD model of target speaker, all adaptation methods show better performance with same amount of adaptation data. Experimental results also reveal that, by fine-tuning together with inserted LN, the performance of the basic only fine-tuning output layer adaptation method, jumps from the pink curve to the blue(OL+LRPD-LN) and green curve(OL+Full-LN). These results indicate that by only fine-tuning the output transformation layer, without tuning more hidden layers, it is not possible to fully adapt to the target voice. This is the case specially when there is more adaptation data available, and the gap between OL and OL+LN method becomes huge as more adaptation data available. Since LRPD-LN and Full-LN are always fine-tuned with OL in our experiments, LRPD-LN and Full-LN is used to indicate OL+LRPD-LN and OL+Full-LN for short.
	
	Moreover, Full-LN is worse than LRPD-LN when adaptation utterances are limited (less than 50). This is because Full-LN introduces too much speaker-specific parameters, and causes over-fitting due to the lack of data. When adaptation utterances are adequate (more than 100), Full-LN is better than LRPD-LN. It is mainly because the number of speaker specific parameters of LRPD-LN is limited with fixed decomposition dimension $r=10$.
	
	\subsubsection{subjective measure}
	\label{subsubsec:subjective_measure_TTS_SAT_xiaoyun_to_xiaoyu}
	\begin{figure}[t]
		\centering
		\centerline{\includegraphics[scale=0.58]{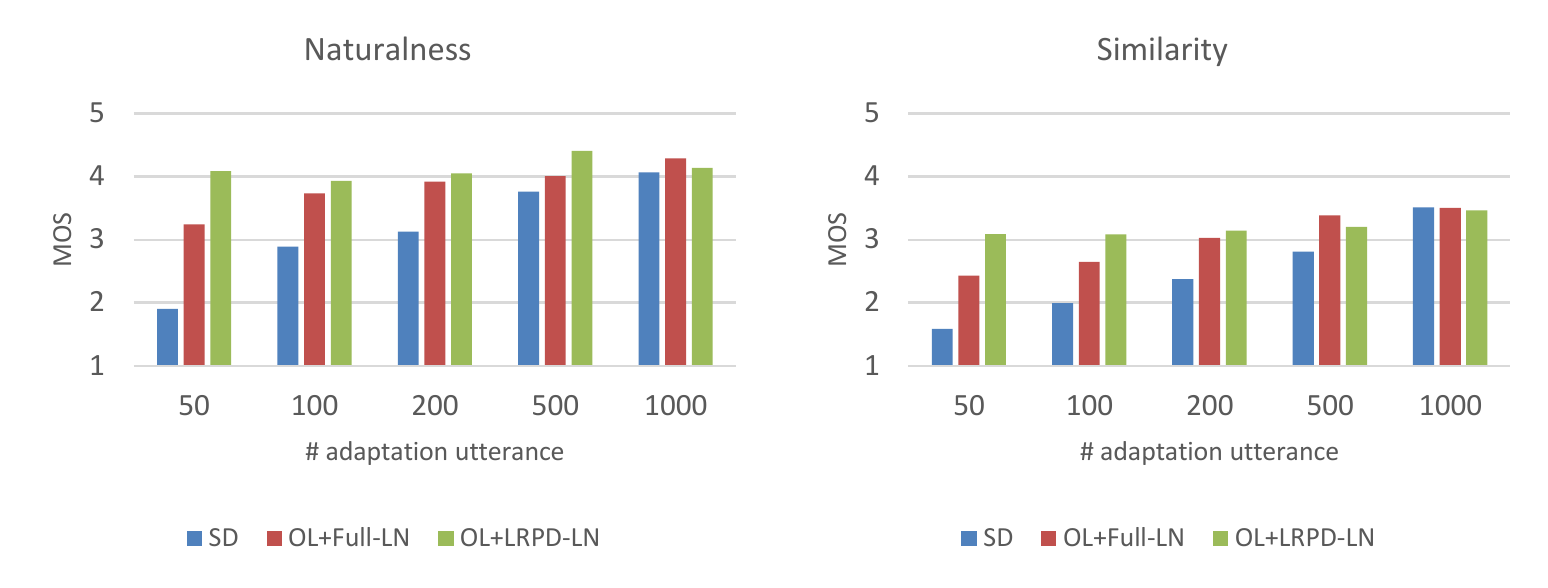}}
		\caption{Subjective tests of testing set utterances (from C-female to B-female).}
		\label{fig:TTS_SAT_xiaoyun_to_xiaoyu_subjective}
	\end{figure}
	
	Naturalness and similarity to target speaker of different systems are shown in Fig. \ref{fig:TTS_SAT_xiaoyun_to_xiaoyu_subjective}. In these two subjective tests, performance of SD (see the blue bar) system degrades quickly when the number of adaptation utterances decrease, and LRPD-LN (see the green bar) based adaptation is more stable compared to SD and Full-LN (see the red bar). Moreover, it's found that both Full-LN and LRPD-LN show better performance than SD with same number of utterances. Full-LN and LRPD-LN with 200 adaptation utterances can achieve similar performance to SD with 1000 utterances. Different to the objective measurements, LRPD-LN outperforms Full-LN until adaptation data reaches 500. This is because the over-fitting makes synthesis voice not stable, sometimes sounds very weird and unintelligible. And we can still draw conclusion that over-fitting makes Full-LN become worse when the adaptation utterances is deficient. 
	
	\subsection{Tough task: from male to female and from female to male}
	\label{subsec:male_to_female}
	Both male to female adaptation and female to male adaptation obtain similar objective trends to Fig.~\ref{fig:TTS_SAT_xiaoyun_to_xiaoyu_objective}, and thus we skip this results and only show the subjective results.
	
	\subsubsection{subjective measure}
	\begin{figure}[t]
		\centering
		\centerline{\includegraphics[scale=0.58]{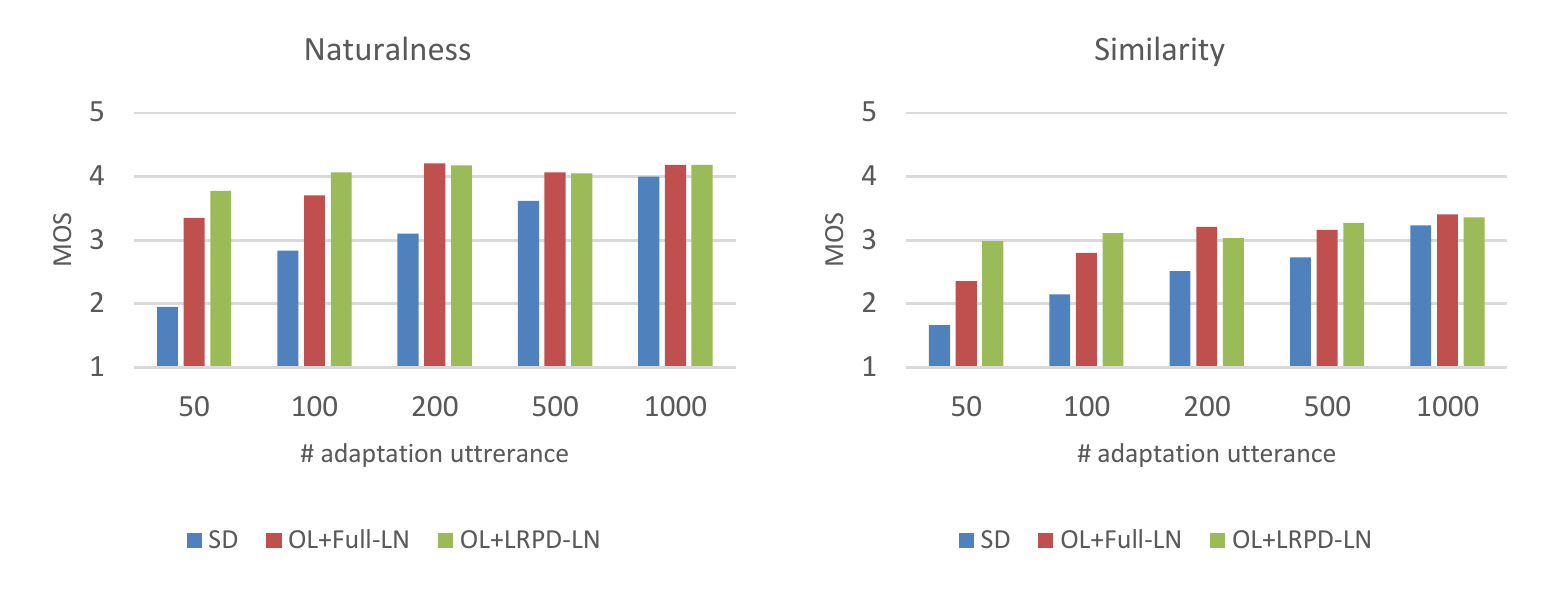}}
		\caption{Subjective tests of testing set utterances (from A-male to B-female).}
		\label{fig:TTS_SAT_xiaogang_to_xiaoyu_subjective}
	\end{figure}
	
	\begin{figure}[t]
		\centering
		\centerline{\includegraphics[scale=0.58]{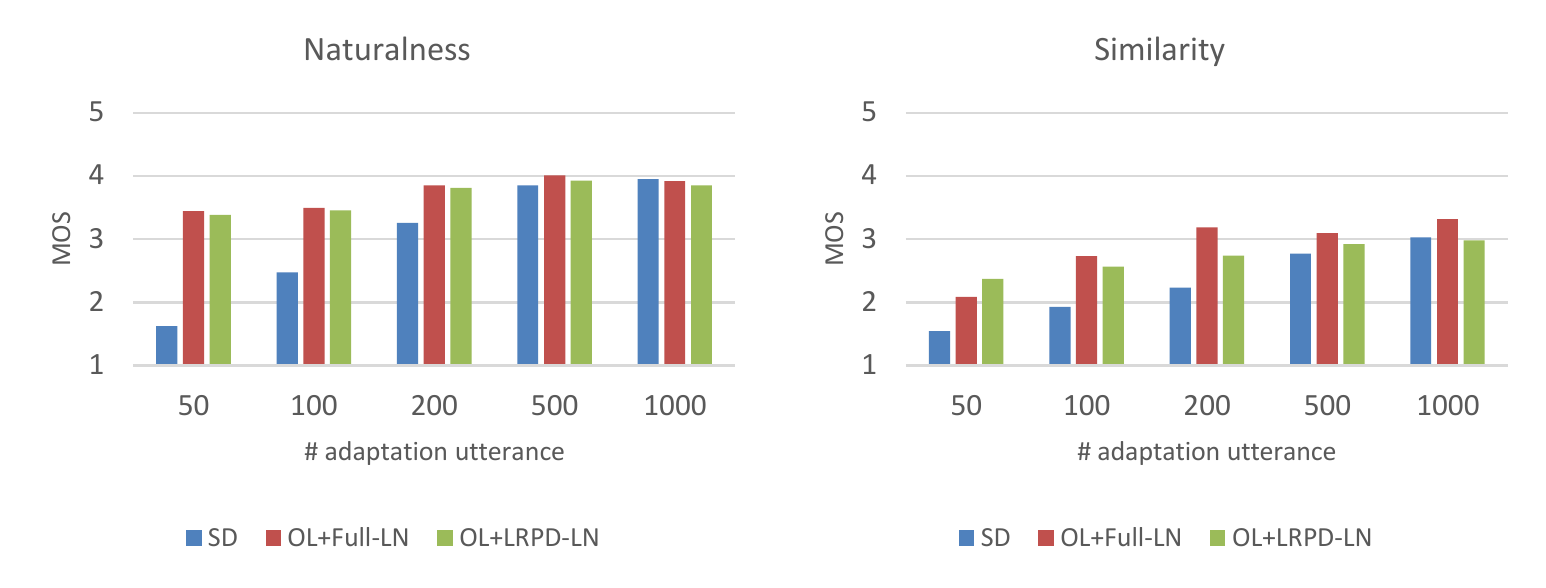}}
		\caption{Subjective tests of testing set utterances (from B-female to A-male).}
		\label{fig:TTS_SAT_xiaoyu_to_xiaogang_subjective}
	\end{figure}
	
	As shown in Fig. \ref{fig:TTS_SAT_xiaogang_to_xiaoyu_subjective} and Fig. \ref{fig:TTS_SAT_xiaoyu_to_xiaogang_subjective}, the same conclusion as in section \ref{subsubsec:subjective_measure_TTS_SAT_xiaoyun_to_xiaoyu} can be made. LRPD-LN is more stable than SD and Full-LN, and the performance of SD declines quickly while decreasing the training utterances number. Also, Full-LN and LRPD-LN based methods both still show better performance than SD with the same utterances number. Similarly, 200 adaptation utterances for Full-LN and LRPD-LN adaptation achieve comparable performance to SD with 1000 utterances. 
	
	In addition, it is interesting to see in Fig. \ref{fig:TTS_SAT_xiaoyu_to_xiaogang_subjective}, the gap between adaptation voices and SD voice are larger when adaptation is extremely small. This may indicate that adaptation works best to if the target speaker is male, and if the adaptation data is limited. However, the gap shrinks quickly as the adaptation data becomes larger. Overall, speaking, Fig. \ref{fig:TTS_SAT_xiaoyu_to_xiaogang_subjective} reveals the same trend as in Fig. \ref{fig:TTS_SAT_xiaoyun_to_xiaoyu_subjective} and Fig. \ref{fig:TTS_SAT_xiaogang_to_xiaoyu_subjective}.

	\section{Conclusion}
	\label{sec:conclusion}
	In this paper, LN based speaker adaptation methods are investigated to speaker adaptation in speech synthesis, and LRPD decomposition for LN is employed to make the adapted voice more stable.  After conducting speaker adaptation from 1) female to female, 2) male to female, and 3) female to male, experimental results show that LN with LRPD decomposition performs most stable when adaptation data is extremely limited. Also, our best SA model with only 200 adaptation utterances achieves comparable quality with SD model trained with 1000 utterances, in both naturalness and similarity to target speaker.

	\bibliographystyle{IEEEbib}
	\bibliography{strings,refs}

\end{document}